\documentclass[useAMS,galley]{mn2e}
\usepackage{graphicx}
\usepackage{times,mathptm}
\newcommand{\ltsima}{$\; \buildrel < \over \sim \;$}
\newcommand{\simlt}{\lower.5ex\hbox{\ltsima}}
\newcommand{\gtsima}{$\; \buildrel > \over \sim \;$}
\newcommand{\simgt}{\lower.5ex\hbox{\gtsima}}

\newcommand{\cgs}{ ${\rm erg~cm}^{-2}~{\rm s}^{-1}$} 
\newcommand{\lum}{\rm erg~s$^{-1}$}

\def\lesssim{\mathrel{\hbox{\rlap{\hbox{\lower4pt\hbox{$\sim$}}}\hbox{$<$}}}}
\def\gtrsim{\mathrel{\hbox{\rlap{\hbox{\lower4pt\hbox{$\sim$}}}\hbox{$>$}}}}

\def\arcmin{\hbox{$^\prime$}}
\def\arcsec{\hbox{$^{\prime\prime}$}}

\def\ab1450{$AB_{1450(1+z)}$}

\def\xray{\hbox{X-ray}}

\def\oiii{\hbox{[O\ {\sc iii}}]}
\def\mgii{\hbox{Mg\ {\sc ii}}}

\def\loiii{$L_{[O\ III]}$}
\newcommand\phn{\phantom{0}}%
\def\chandra{{\it Chandra\/}}

\def\heao1{{\it HEAO-1\/}}

\def\rosat{{\it ROSAT\/}}

\def\xmm{{XMM-{\it Newton\/}}}

\def\aj{AJ}
\def\araa{ARA\&A}
\def\apj{ApJ}

\def\apjs{ApJS}

\def\aap{A\&A}

\def\mnras{MNRAS}

\def\procspie{Proc.~SPIE}

\title[THE CHANDRA QUEST FOR TYPE~2 QUASARS]
{The quest for Type~2 quasars: Chandra observations of luminous obscured quasars in the 
Sloan Digital Sky Survey}
\author[C. Vignali et al.]
{
Cristian Vignali,$^{1,2}$\thanks{E-mail: cristian.vignali@oabo.inaf.it (CV); 
d.m.alexander@dur.ac.uk (DMA); andrea.comastri@oabo.inaf.it (AC).}
Dave M. Alexander$^{3}$\footnotemark[1] and 
Andrea Comastri$^{2}$\footnotemark[1] \\ \\ 
$^{1}$ Dipartimento di Astronomia, Universit\`a degli Studi di Bologna, 
Via Ranzani 1, 40127 Bologna, Italy \\
$^{2}$ INAF -- Osservatorio Astronomico di Bologna, Via Ranzani 1, 
40127 Bologna, Italy \\
$^{3}$ Department of Physics, Durham University, South Road, Durham DH1~3LE \\
}

\begin{document}

\date{Accepted 2006 September 2. Received 2006 September 1; in original form 2006 July 6}

\pagerange{\pageref{firstpage}--\pageref{lastpage}} \pubyear{2006}

\maketitle

\label{firstpage}

\begin{abstract}
We report on new \chandra\ exploratory observations of six candidate Type~2 
quasars at \hbox{$z$=0.49--0.73} selected among the most \oiii\ luminous emitters 
from the Sloan Digital Sky Survey (SDSS). 
Under the assumption that \oiii\ is a proxy for the intrinsic 
luminosity of the central source, their predicted rest-frame \xray\ luminosities are 
$L_{2-10keV}\approx10^{45}$~\lum. 
For two of the targets, the photon statistics are good enough to allow for basic \xray\ 
spectral analyses, which indicate the presence of intrinsic absorption 
($\approx$~10$^{22-23}$~cm$^{-2}$) and luminous \xray\ emission 
($L_{\rm X}\gtrsim10^{44}$~\lum). 
Of the remaining four targets, two are detected with only a few (3--6) \xray\ counts, 
and two are undetected by \chandra. 
If these four sources have the large intrinsic \xray\ luminosities predicted by 
the \oiii\ emission, then their nuclei must be heavily obscured 
($N_{\rm H}>$~few~$\times10^{23}$~cm$^{-2}$) and some might be Compton thick 
($N_{\rm H}>1.5\times10^{24}$~cm$^{-2}$). 
We also present the results for two Type~2 quasar candidates serendipitously lying in the 
fields of the \chandra\ targets, and provide an up-to-date compilation of the 
\xray\ properties of eight additional SDSS Type~2 quasars from archival \chandra\ 
and \xmm\ observations (five with moderate-quality \xray\ data). 
The combined sample of 16 SDSS Type~2 quasars (10 \xray\ detections) provides 
further evidence that a considerable fraction of optically selected 
Type~2 quasars are obscured in the \xray\ band 
(at least all the objects with moderate-quality \xray\ spectra), 
lending further support to the findings 
presented in Vignali, Alexander and Comastri (2004a) and unification schemes 
of Active Galactic Nuclei, and confirms the reliability of \oiii\ emission in predicting 
the \xray\ emission in obscured quasars. 
\end{abstract}


\section{Introduction}
The quest for the identification of Type~2 quasars, the high-luminosity, long-sought after 
``big cousins" of local Seyfert~2 galaxies (i.e., characterized by high-ionization, 
narrow emission lines and a lack of broad emission lines in the rest-frame optical/ultraviolet 
spectra), has been the topic of numerous investigations over the past few years. 
Although there is not a unique definition of Type~2 quasars in the \xray\ band, generally the most 
accepted view is that these sources should be the luminous and obscured 
($L_{\rm X}>10^{44}$~\lum\ and $N_{\rm H}>10^{22}$~cm$^{-2}$) Active Galactic Nuclei (AGN) 
population predicted by unification schemes of AGN (e.g., Antonucci 1993) and required by 
many synthesis models of the \xray\ background (XRB; e.g., Comastri et al. 2001; 
Gilli, Salvati \& Hasinger 2001). 
For example, in the most recent version of the synthesis model of Gilli, Comastri \& Hasinger 
(2006), the contribution of Type~2 quasars to the XRB is estimated to be 
$\approx$~15~per cent. 

Over the past 4--5 years, deep \xray\ surveys 
(such as those conducted in the \chandra\ deep fields; e.g., Giacconi et al. 
2002; Alexander et al. 2003; see Brandt \& Hasinger 2005 for a recent review) 
performed with \chandra\ and \xmm\ have resolved a large fraction 
of the XRB ($\approx$~80~per~cent in the 2--10~keV band; 
e.g., Bauer et al. 2004; Hickox \& Markevitch 2006; but see also Worsley et al. 2005) 
and detected many high-redshift Type~2 quasar candidates (e.g., 
Crawford et al. 2001, 2002; Mainieri et al. 2002; Barger et al. 2003). 
Optical and infrared spectroscopy is critical to determine the redshifts of 
these sources and confirm that these candidates are ``genuine'' 
Type~2 quasars (i.e., matching both the optical and \xray\ definitions). 
However, the majority of these objects are optically faint (most likely because they lie at high 
redshifts) and are therefore challenging targets to obtain spectroscopic redshifts. 
As a consequence, although considerable efforts have been made, 
the number of spectroscopically confirmed Type~2 quasars from these surveys is comparatively small 
(e.g., Norman et al. 2002; Stern et al. 2002; 
Fiore et al. 2003; Perola et al. 2004; Gandhi et al. 2004; 
Caccianiga et al. 2004; Szokoly et al. 2004; Mainieri et al. 2005a; 
Maiolino et al. 2006), 
despite the fact that as many as 15--20~per~cent of luminous obscured quasars are 
predicted to be observed in the deepest \xray\ surveys 
(Alexander et al. 2001; Mainieri et al. 2005b).

The goal of our project is to select optically luminous 
Type~2 quasar candidates identified on the basis of high 
\oiii5007\AA\ luminosity from the ground-based, large-area Sloan Digital Sky 
Survey (SDSS; York et al. 2000) spectroscopic data 
and then follow up these sources with \xray\ observations to confirm whether or not 
they are luminous and obscured also in the \xray\ band. 
The advantage of this approach, compared to the study of Type~2 quasars 
found in moderately deep and ultra-deep \xray\ surveys, 
is that at the optical magnitude limit of the SDSS, Type~2 quasars are generally easy to study at 
both optical and \xray\ wavelengths. In addition, 
\xray\ studies of optically selected Type~2 quasars 
can explore whether selection in the optical and \xray\ bands 
identifies sources with intrinsically different properties. 
%
%
Due to observational limitations (e.g., small-size samples, low percentages of spectroscopic 
identifications, limited regions of the sky being surveyed), 
such studies have been challenging or impossible in the past. 

In this paper we extend the work presented by 
Vignali, Alexander and Comastri (2004a, hereafter V04; 2004b) and focus 
on the study of the \xray\ properties of Type~2 quasar candidates originally 
selected from the SDSS in the redshift range 
$\approx$~0.3--0.8. These sources are classified on the basis of high-excitation, 
narrow emission lines without underlying broad components and with line ratios 
characteristic of non-stellar ionizing radiation (Zakamska et al. 2003; see 
Zakamska et al. 2004, 2005 and 2006 for observations of these objects 
at other wavelengths). 
From the analysis of archival \rosat\ and \xmm\ observations, 
V04 were able to place constraints in a statistical sense on the \xray\ emission 
of 17 of these objects (three \xray\ detections and 14 upper limits). 
Assuming the correlation between the \oiii\ and the \hbox{2--10~keV} flux 
found for Seyfert~2 galaxies (Mulchaey et al. 1994) and adopting the parameterization 
reported in $\S$3.2 of Collinge \& Brandt (2000), V04 
found that at least 47~per~cent of the observed sample showed indications of 
\xray\ absorption, including the four highest luminosity sources with 
predicted intrinsic luminosities of $\approx$~10$^{45}$~\lum. 

Here we present the exploratory \chandra\ observations for six 
additional SDSS Type~2 quasar candidates and provide an up-to-date list 
of the \xray\ properties of Type~2 quasar candidates for which archival 
\chandra\ and \xmm\ data are available (see also Ptak et al. 2006, 
hereafter P06, where pointed and archival \xray\ observations of a sample of mostly bright 
SDSS Type~2 quasar candidates are presented). In total we have 16 sources (10 \xray\ 
detections) with available \chandra\ and \xmm\ data at present. 
The choice of \chandra\ and \xmm\ is motivated by their broad-band coverage 
(thus minimizing the uncertainties due to extrapolations as in the case of 
\rosat\ data), relatively large field-of-view, spectroscopic capabilities, 
and good sensitivity to relatively \xray\ faint sources. 

Hereafter we adopt the ``concordance'' (WMAP) cosmology 
($H_{0}$=70~km~s$^{-1}$~Mpc$^{-1}$, $\Omega_{\rm M}$=0.3, and 
$\Omega_{\Lambda}$=0.7; Spergel et al. 2003).

\section{The main sample: Chandra Cycle~6 observations}

\subsection{The sample selection}
\begin{figure}
\includegraphics[angle=0,width=85mm]{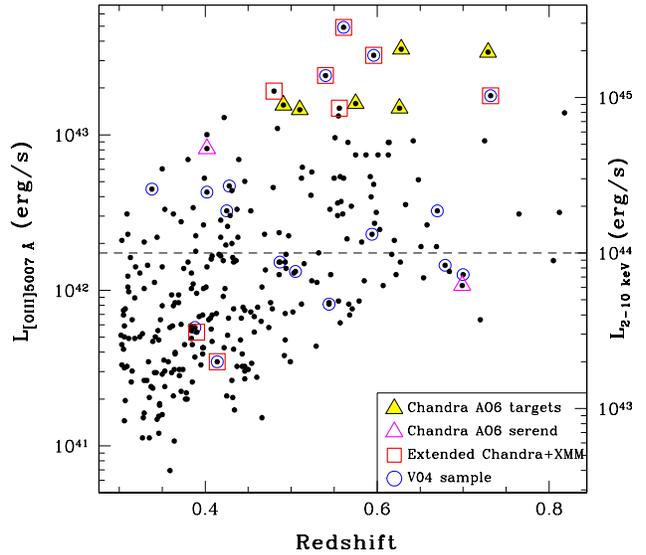}
\caption{
Logarithm of the measured \loiii\ luminosities 
vs. redshift for all of the sources in the Zakamska et al. (2003) 
catalog (small filled circles). 
The y-axis on the right-hand side of the plot indicates the predicted 
\hbox{2--10~keV} luminosity on the basis of the \oiii\ emission (see $\S$2.1). 
The dashed line provides an approximate separation between the locus of 
high-luminosity objects (quasar candidates, above the line) from that presumably 
populated by Seyfert galaxies (below the line); see Zakamska et al. (2003) for further 
details. 
The inset provides a key description of the \xray\ observations. 
}
\label{lxoiiiz}
\end{figure}

In the study conducted by V04, which used primarily \rosat\ observations, 
$\approx$~40~per~cent of the Type~2 quasar candidates had predicted intrinsic 
\hbox{2--10~keV} luminosities 
closer to those typically found for Seyfert galaxies than for quasars 
(open circles in Fig.~\ref{lxoiiiz}; also see the summary of the results in Table~2 of V04). 
The principal goal of the work described here is to extend the study of V04 
with the much higher sensitivity of \chandra. 
We selected six candidate Type~2 quasars on the basis of their luminous 
\oiii\ emission which, in principle, should identify the most \xray\ luminous 
members of the Type~2 quasar population (see filled triangles in Fig.~\ref{lxoiiiz}). 
Even assuming the lowest \xray\ luminosities predicted from 
the dispersion in the $L_{[O\ III]}$--$L_{\rm 2-10~keV}$ 
correlation, the six targets should still easily lie in the quasar regime 
($L_{\rm 2-10~keV}>10^{44}$~erg~s$^{-1}$). 
Hereafter we refer to this sample as the ``main'' \chandra\ sample. 
%

\subsection{Chandra data reduction and analysis}
The six SDSS Type~2 quasar candidates were targeted by \chandra\ during 
Cycle~6 with the Advanced CCD Imaging Spectrometer (ACIS; 
Garmire et al. 2003) and the S3 CCD at the aimpoint, using short 
($\approx$~6.3--11.5~ks) exposures. Standard data reduction procedures were adopted 
(see $\S$2 of Vignali et al. 2005 for a detailed description) using the 
\chandra\ Interactive Analysis of Observations ({\sc ciao}) Version~3.2 
software. 
%
\begin{figure*}
\includegraphics[angle=0,width=170mm]{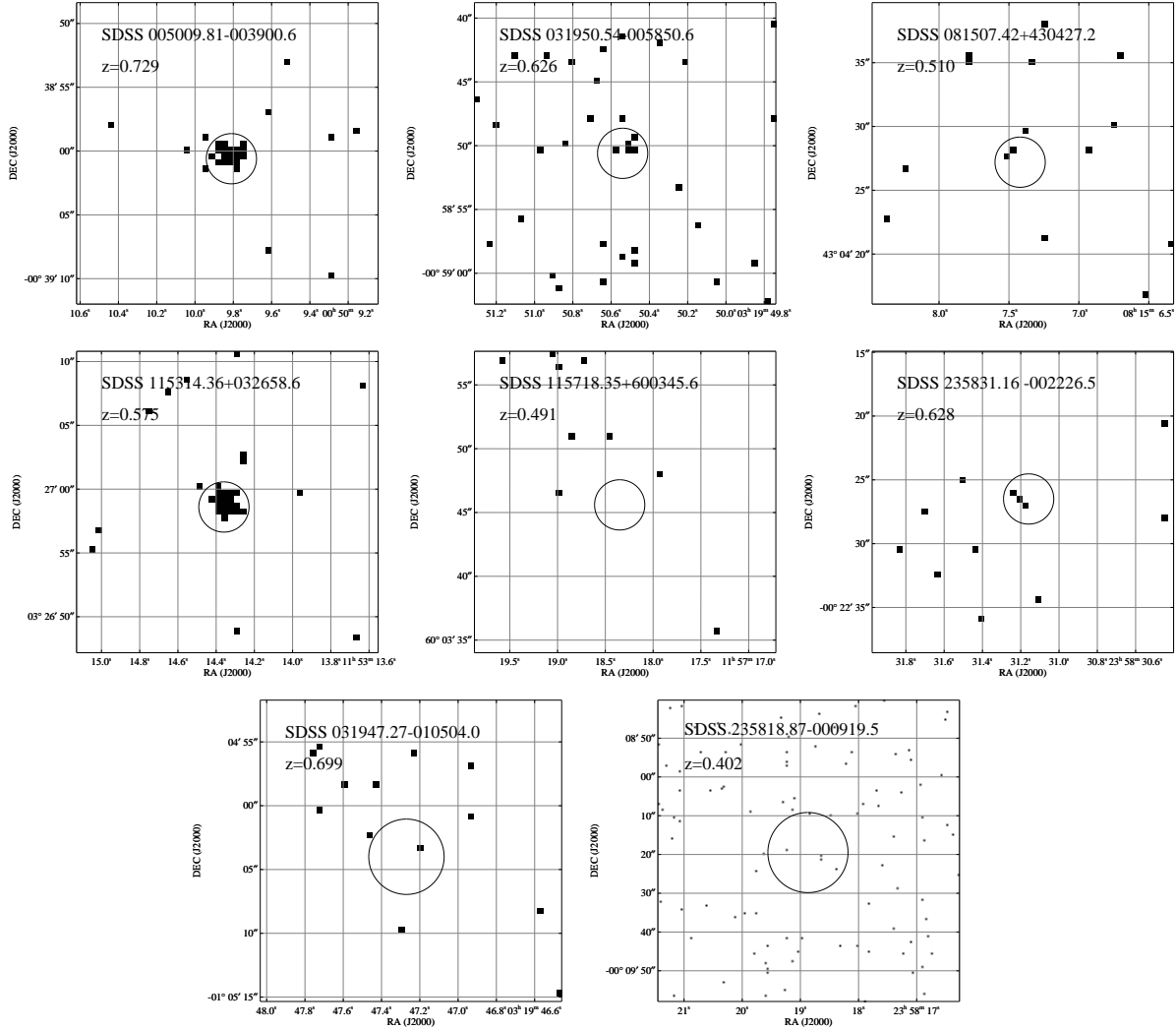}
\vglue-6cm
\caption{
Full-band (0.5--8~keV) raw images of the quasars observed by \chandra\ in 
Cycle~6 and presented in this paper; the top six are the main targets and 
the bottom two are the serendipitous sources. 
The panels are \hbox{25\arcsec $\times$ 25\arcsec} for all of the images,  
except for SDSS~J2358$-$0009 ($\approx$~80\arcsec $\times$ 80\arcsec); 
North is up, and East to the left. 
For the six main targets, \hbox{2\arcsec--radius} circles around the 
SDSS positions of the quasars are shown, while for the last two
sources, which serendipitously fell within the \chandra\
field-of-view, the \hbox{3\arcsec--radius} (SDSS~J0319$-$0105) and 
\hbox{10\arcsec--radius} (SDSS~J2358$-$0009) source extraction regions 
are shown (see text for details). 
These two sources, along with two of the main targets, are not 
detected in the current \chandra\ observations. 
}
\label{chandra_images}
\end{figure*}

Source detection was carried out with {\sc wavdetect} (Freeman et al. 2002) 
using a false-positive probability threshold of 10$^{-6}$ except for the faintest sources, 
where a false-positive probability threshold of 10$^{-4}$ was adopted. 
The observation log is shown in Table~\ref{obs_log}; hereafter we 
will refer to the SDSS sources with their abbreviated names. 
For one source (SDSS~J0319$-$0058), a high-background interval was removed 
from the \chandra\ observation, yielding a final useful exposure of 11.10~ks. 

Four sources were detected with $\approx$~3--80 net counts in the observed 
\hbox{0.5--8~keV} (full-band) energy range (see Table~\ref{chandra_counts}); 
their \xray\ position is within $\approx$~0.8\arcsec\ of 
the optical position (see Table~\ref{obs_log}), as expected given the 
positional accuracy of \chandra\ ACIS for on-axis observations. 
The 3-count source (SDSS~J2358$-$0022)
was detected using a false-positive probability threshold of 
10$^{-4}$; given the small number of pixels being searched due to the known 
source position and the sub-arcsec on-axis angular resolution of \chandra, 
the probability that this detection is spurious is low (2.8$\times10^{-4}$). 
One target (SDSS~J0815$+$4304) has 2 counts within a 2\arcsec-radius circle, 
but neither {\sc wavdetect} nor Monte-Carlo simulations (giving a $\approx$~3.5$\sigma$ 
significance for this source being actually detected; 
see Vignali et al. 2005 for details on these simulations) provide conclusive support to the \xray\ 
detection of this source. 
%
\begin{table*}
\centering
\begin{minipage}{160mm}
\caption{SDSS Type~2 quasar candidates: \chandra\ and \xmm\ observation logs.}
\label{obs_log}
\scriptsize
\begin{tabular}{rccccccccc}
\hline
Src. ID \# & Object Name &  & X-ray & X-ray & $\Delta_{\rm Opt-X}$ & Obs.~Date &
Exp.~Time & Off-axis & Seq. Num. \\
  &  & $z$ & ($\alpha_{2000}$) & ($\delta_{2000}$) & (arcsec) & 
   & (ks) & (arcmin) & \\
\hline

\multicolumn{10}{c}{\bf Sources observed as targets in Chandra AO6} \\ \\
  7 & SDSS~J005009.81$-$003900.6 & 0.729 & 00 50 09.8 & $-$00 39 00.5 & 0.3      & 2005 Aug 28     & {\phn}7.85         & {\phn}0.0 & 701116       \\
 83 & SDSS~J031950.54$-$005850.6 & 0.626 & 03 19 50.5 & $-$00 58 50.5 & 0.3      & 2005 Mar 10     &      11.10$^{a}$   & {\phn}0.0 & 701117       \\
119 & SDSS~J081507.42$+$430427.2 & 0.510 & \dotfill   & \dotfill      & \dotfill & 2004 Dec 25     & {\phn}8.17         & {\phn}0.0 & 701118       \\ 
196 & SDSS~J115314.36$+$032658.6 & 0.575 & 11 53 14.4 & $+$03 26 58.6 & 0.3      & 2005 Apr 10--11 & {\phn}8.17         & {\phn}0.0 & 701119       \\
197 & SDSS~J115718.35$+$600345.6 & 0.491 & \dotfill   & \dotfill      & \dotfill & 2005 Jun 03     & {\phn}6.97         & {\phn}0.0 & 701120       \\
290 & SDSS~J235831.16$-$002226.5
                                 & 0.628 & 23 58 31.2 & $-$00 22 26.8 & 0.8      & 2005 Aug 08     & {\phn}6.27         & {\phn}0.0 & 701121       \\
\hline 
\multicolumn{10}{c}{\sf Serendipitous sources in Chandra AO6 observations} \\ \\
 82 & SDSS~J031947.27$-$010504.0 & 0.699 & \dotfill   & \dotfill      & \dotfill & 2005 Mar 10     & {\phn}9.67$^{a,b}$ & {\phn}7.4 & 701117       \\ 
289 & SDSS~J235818.87$-$000919.5 & 0.402 & \dotfill   & \dotfill      & \dotfill & 2005 Aug 08     & {\phn}5.79$^{b}$   &      13.3 & 701121       \\ 
\hline 
\multicolumn{10}{c}{\sf Archival \chandra\ observations [targets$+$serendipitous sources]} \\ \\
 15 & SDSS~J011522.19$+$001518.5 & 0.390 & 01 15 22.2 & $+$00 15 18.8 & 0.4      & 2002 Nov 01     &      33.70$^{b}$   & {\phn}2.1 & 800204 \\ 
113 & SDSS~J080154.24$+$441234.0 & 0.556 & 08 01 54.3 & $+$44 12 33.8 & 0.4      & 2003 Nov 27     & {\phn}9.82         & {\phn}0.0 & 701011 \\ 
130 & SDSS~J084234.94$+$362503.1 & 0.561 & \dotfill   & \dotfill      & \dotfill & 1999 Oct 21     & {\phn}6.97$^{b}$   & {\phn}7.5 & 800040 \\ 
207 & SDSS~J123215.81$+$020610.0 & 0.480 & 12 32 15.8 & $+$02 06 10.4 & 0.4      & 2005 Apr 21     & {\phn}9.61         & {\phn}0.0 & 700992 \\ 
\hline 
\multicolumn{10}{c}{\sf Archival \xmm\ observations [targets$+$serendipitous sources]} \\ \\
 34 & SDSS~J021047.01$-$100152.9 & 0.540 & 02 10 47.1 & $-$10 01 54.1 & 1.7      & 2004 Jan 12--13 & 7.98/11.08/11.30$^{a}$   & {\phn}0.0 & 0204340201   \\ 
 59 & SDSS~J024309.79$+$000640.3 & 0.414 & \dotfill   & \dotfill      & \dotfill & 2000 Jul 29     &14.89/19.44/16.93$^{a,b}$ &10.4 & 0111200101   \\ 
204 & SDSS~J122656.48$+$013124.3$\ddagger$ 
                                 & 0.732 & 12 26 56.5 & $+$01 31 24.3 & 1.6      & 2001 Jun 23     & 6.17/5.49/7.37$^{a,b}$   & {\phn}6.0 & 0110990201   \\ 
256 & SDSS~J164131.73$+$385840.9 & 0.596 & 16 41 31.7 & $+$38 58 39.8 & 1.0      & 2004 Aug 20     & 12.20/16.51/16.61$^{a}$  & {\phn}0.0 & 0204340101
  \\ 
\hline
\end{tabular}
\end{minipage}
\begin{minipage}{155mm}
The source ID is taken from Table~1 of Zakamska et al. (2003). 
The optical positions of the quasars can be drawn from their 
SDSS name, while the \xray\ positions for the \xray\ detected sources 
have been obtained from {\sc wavdetect} using the full-band image. 
All of the exposure times were corrected for detector dead time. 
For the \xmm\ observations, all the exposure times of pn, MOS1, and MOS2 
detectors are shown (after cleaning for the high-background periods). 
%
Results from archival \chandra\ and \xmm\ observations were also reported in 
P06. 
$^{a}$ Corrected for high-background periods. 
$^{b}$ The exposure time has been obtained from the exposure map, since the 
source is not located on axis. 
$\ddagger$ Already presented by Vignali, Alexander and Comastri (2004). 
\end{minipage}
\end{table*}

%

Two additional SDSS Type~2 quasar candidates fell serendipitously in the field-of-view of 
the current \chandra\ observations (SDSS~J0319$-$0105 and SDSS~J2358$-$0009, 
plotted as open triangles in Fig.~\ref{lxoiiiz}), 
at large off-axis angles (7.4\arcmin\ and 13.3\arcmin, respectively; 
see Table~\ref{obs_log} for further details). 
None of them was detected with {\sc wavdetect} up to false-positive probability thresholds of 
10$^{-3}$. The photometric results for these two sources 
[adopting source extraction circles with radii of 3\arcsec\ and 10\arcsec, 
respectively, to account for the broadening of the point spread function 
(PSF) at their large off-axis angles] are shown in Table~\ref{chandra_counts}. 

The full-band raw (un-smoothed) images of the six main targets and two 
serendipitous sources are shown in Fig.~\ref{chandra_images}. 
%
\begin{table}
\begin{minipage}{90mm}
\caption{X-ray counts: \chandra\ observations}
\label{chandra_counts}
\scriptsize
\begin{tabular}{lcrrrr}
\hline
   & & \multicolumn{3}{c}{X-ray counts} & Count~rate \\
\cline{3-5} \\
Source & Sample & [0.5--2 keV] & [2--8 keV] & [0.5--8 keV] & [0.5--8 keV] \\
SDSS~J0050$-$0039 & M & 2.8$^{+2.9}_{-1.6}$        & 37.1$^{+7.2}_{-6.1}$    & 40.7$^{+7.4}_{-6.4}$    & 5.18$^{+0.94}_{-0.81}$ \\ 
SDSS~J0319$-$0058 & M & 2.0$^{+2.7}_{-1.3}\dagger$ & 2.8$^{+2.9}_{-1.6}$     & 6.0$^{+3.6}_{-2.4}$     & 0.54$^{+0.32}_{-0.22}$ \\ 
SDSS~J0815$+$4304 & M & $<4.8$                     & $<4.8$                  & $<6.4\ast$              & $<0.78$                \\ 
SDSS~J1153$+$0326 & M & 38.4$^{+7.3}_{-6.2}$       & 42.2$^{+7.6}_{-6.5}$    & 80.4$^{+10.1}_{-9.0}$   & 9.84$^{+1.26}_{-1.10}$ \\
SDSS~J1157$+$6003 & M & $<3.0$                     & $<3.0$                  & $<3.0$                  & $<0.43$                \\ 
SDSS~J2358$-$0022 & M & $<4.8$                     & $<6.4$$\ddagger$        & 2.9$^{+2.9}_{-1.6}$     & 0.46$^{+0.46}_{-0.26}$ \\ 
\hline
SDSS~J0319$-$0105 & S & $<3.0$                     & $<4.4$                  & $<4.4$                  & $<0.50$                \\ 
SDSS~J2358$-$0009 & S & $<4.1$                     & $<5.9$                  & $<6.3$                  & $<1.09$                \\
\hline
SDSS~J0115$+$0015 & A & 115.7$^{+11.8}_{-10.7}$    & 202.9$^{+15.3}_{-14.2}$ & 317.7$^{+18.9}_{-17.8}$ & 9.42$^{+0.56}_{-0.52}$ \\  
SDSS~J0801$+$4412 & A & 4.9$^{+3.4}_{-2.1}$        & 33.3$^{+6.8}_{-5.7}$    & 38.1$^{+7.2}_{-6.1}$    & 3.88$^{+0.73}_{-0.62}$ \\ 
SDSS~J0842$+$3625 & A & $<9.9$                     & $<6.6$                  & $<7.7$                  & $<1.11$                \\
SDSS~J1232$+$0206 & A & 2.0$^{+2.7}_{-1.3}\dagger$ & 3.9$^{+3.2}_{-1.9}$     & 5.9$^{+3.6}_{-2.4}$     & 0.61$^{+0.37}_{-0.25}$ \\
\hline
\end{tabular}
\end{minipage}
\begin{minipage}{87mm} 
``M'' refers to the main target sample, ``S'' to the serendipitous sources 
and ``A'' to the archival sources. 
Errors on the \xray\ net counts (i.e., background-subtracted) 
were computed according to Gehrels (1986). 
The upper limits are at the 95\% confidence level and were computed 
according to Kraft, Burrows \& Nousek (1991). 
Count rates  are in units of $10^{-3}$ counts~s$^{-1}$. \\
$\ast$ The two \xray\ counts in the source extraction ragion (see Fig.~2) 
are contiguous but neither {\sc wavdetect} nor Monte-Carlo 
simulations provide conclusive support to the \xray\ detection of this source, therefore 
an upper limit is reported according to Kraft et al. (1991). 
$\dagger$ The two \xray\ counts are contiguous. 
$\ddagger$ The two \xray\ counts are not contiguous, therefore an upper limit is reported 
according to Kraft et al. (1991). 
\end{minipage}
\end{table}

%
\begin{table*}
\centering
\begin{minipage}{155mm}
\caption{Properties of the SDSS Type~2 quasars observed by \chandra\ and \xmm.}
\label{xray_param}
\begin{tabular}{cccccccc}
\hline
Object & $N_{{\rm H}_{\rm gal}}$ & $\log~L_{[OIII]}$ & $F_{2-10~keV}$
& $N_{{\rm H}_{\rm z}}$ & $L_{2-10~keV}$ & $L_{2-10~keV}$ (pr.) & $S_{1.4~GHz}$ \\
(1) & (2) & (3) & (4) & (5) & (6) & (7) & (8) \\
\hline
SDSS~J0050$-$0039 & 2.57 &  9.94 & 1.6$\times10^{-13}$  & 3.76$^{+2.38}_{-1.58}\times10^{23}$ & 7.2$\times10^{44}$  & 4.7$\times10^{44}$--8.1$\times10^{45}$ & 4.32     \\ 
SDSS~J0319$-$0058 & 6.05 &  9.58 & 2.2$\times10^{-15}$  & \dotfill                            & 3.7$\times10^{42}$  & 2.0$\times10^{44}$--3.6$\times10^{45}$ & $<0.143$ \\
SDSS~J0815$+$4304$\ast$
                  & 5.03 &  9.57 & $<3.1\times10^{-15}$ & \dotfill                            & $<3.2\times10^{42}$ & 2.0$\times10^{44}$--3.5$\times10^{45}$ & 6.09     \\ 
SDSS~J1153$+$0326 & 1.89 &  9.61 & 8.2$\times10^{-14}$  & 1.54$^{+0.90}_{-0.54}\times10^{22}$ & 1.2$\times10^{44}$  & 2.2$\times10^{44}$--3.8$\times10^{45}$ & 2.09     \\ 
SDSS~J1157$+$6003 & 1.65 &  9.60 & $<1.6\times10^{-15}$ & \dotfill                            & $<1.5\times10^{42}$ & 2.1$\times10^{44}$--3.7$\times10^{45}$ & 1.54     \\ 
SDSS~J2358$-$0022 & 3.29 &  9.96 & 1.9$\times10^{-15}$  & \dotfill                            & 3.2$\times10^{42}$  & 4.9$\times10^{44}$--8.5$\times10^{45}$ & $<0.143$ \\
\hline
SDSS~J0319$-$0105 & 5.92 &  8.44 & $<2.9\times10^{-15}$ & \dotfill                            & $<6.4\times10^{42}$ & 1.5$\times10^{43}$--2.6$\times10^{44}$ & $<0.137$ \\
SDSS~J2358$-$0009 & 3.25 &  9.32 & $<5.4\times10^{-15}$ & \dotfill                            & $<3.1\times10^{42}$ & 1.1$\times10^{44}$--2.0$\times10^{45}$ & $<0.134$ \\
\hline
SDSS~J0115$+$0015 & 3.42 &  8.14 & 1.3$\times10^{-13}$  & 3.25$^{+0.64}_{-0.56}\times10^{22}$ & 8.1$\times10^{43}$  & 7.4$\times10^{42}$--1.3$\times10^{44}$ & $<0.150$ \\ 
SDSS~J0801$+$4412 & 4.79 &  9.58 & 1.4$\times10^{-13}$  & 4.29$^{+1.90}_{-2.10}\times10^{23}$ & 4.2$\times10^{44}$  & 2.0$\times10^{44}$--3.6$\times10^{45}$ & $<0.130$ \\
SDSS~J0842$+$3625$\ast$ 
                  & 3.41 & 10.10 & $<5.9\times10^{-15}$ & \dotfill                            & $<7.7\times10^{42}$ & 6.8$\times10^{44}$--1.2$\times10^{46}$ & 2.16     \\ 
SDSS~J1232$+$0206 & 1.80 &  9.69 & 2.3$\times10^{-15}$  & \dotfill                            & 2.0$\times10^{42}$  & 2.6$\times10^{44}$--4.6$\times10^{45}$ & $<1.117$ \\ 
\hline
SDSS~J0210$-$1001 & 2.17 &  9.79 & 1.6$\times10^{-13}$  & 8.13$^{+2.39}_{-1.92}\times10^{22}$ & 2.2$\times10^{44}$  & 3.3$\times10^{44}$--5.8$\times10^{45}$ & $<0.157$ \\
SDSS~J0243$+$0006 & 3.56 &  7.95 & $<3.6\times10^{-15}$ & \dotfill                            & $<2.2\times10^{42}$ & 4.8$\times10^{42}$--8.3$\times10^{43}$ & $<0.276$ \\
SDSS~J1226$+$0131 & 1.84 &  9.66 & 1.7$\times10^{-13}$  & 2.63$^{+0.59}_{-0.47}\times10^{22}$ & 4.3$\times10^{44}$  & 2.5$\times10^{44}$--4.3$\times10^{45}$ & $<0.742$ \\
SDSS~J1641$+$3858$\ast$~$\dagger$ 
                  & 1.16 &  9.92 & 4.5$\times10^{-13}$  & 5.51$^{+0.48}_{-0.61}\times10^{22}$ & 7.4$\times10^{44}$  & 4.5$\times10^{44}$--7.8$\times10^{45}$ & 3.54     \\ 
\hline
\end{tabular}
\end{minipage}
\begin{minipage}{155mm}
(1) Abbreviated SDSS name; 
(2) Galactic column density, from Dickey \& Lockman (1990), in units of $10^{20}$~cm$^{-2}$; 
(3) log of the [OIII] line luminosity, in units of $L_{\odot}$ (from Zakamska et al. 2003); 
(4) Galactic absorption-corrected flux (or upper limit) in the 2--10~keV band, either extrapolated from the observed 
    \hbox{0.5--8~keV} (\hbox{0.5--10~keV}) count rate or upper limit in \chandra\ (\xmm) observations, 
    assuming a power law with $\Gamma=2.0$ (typical for the AGN \xray\ emission) or obtained 
    directly from the \xray\ spectral fitting (when possible) with $\Gamma$ frozen to 2.0, for consistency with the other sources. 
    The flux is in units of erg~cm$^{-2}$~s$^{-1}$; 
(5) intrinsic column density, obtained from the \xray\ spectral fitting and using a power law with $\Gamma=2.0$, 
    in units of cm$^{-2}$; 
(6) 2--10~keV rest-frame luminosity, obtained through the observed flux (or upper limit), 
    corrected for the effect of absorption (when the column density can be measured directly from the spectral fit), 
    in units of erg~s$^{-1}$; 
(7) 2--10~keV luminosity range predicted from the \oiii\ line vs. hard \xray\ flux correlation 
    (Mulchaey et al. 1994; see also $\S$2.1), in units of erg~s$^{-1}$; 
(8) radio flux density (or 1$\sigma$ upper limits) at 1.4~GHz from the FIRST (Becker, White \& Helfand 1995), in units of mJy.  
    For SDSS~J0050$-$0039 and SDSS~J0815$+$4304, the radio flux densities reported here are the integrated values, 
    while for the remaining radio-detected sources, the peak values are shown (since there is no evidence for extension in the radio band).  
We note that constraints on further 12 Type~2 quasar candidates (2 of which \xray\ detected) are presented by V04. \\
$\ast$ SDSS~J0815$+$4304, SDSS~J0842$+$3625 and SDSS~J1641$+$3858 have ``mean polarization'' values (averaged 
over the observed \hbox{2820--6530\AA} band and uncorrected for the low dilution by the host galaxy) 
of 5.0, 15.4, and 4.6~per cent, respectively (Zakamska et al. 2005). 
In particular, spectro-polarimetry shows the presence of a broad H$\gamma$ emission line in 
SDSS~J0842$+$3625 and SDSS~J1641$+$3858 ($FWHM\approx$~4900 and 3200~km~s$^{-1}$, respectively) and narrow \mgii\ 
and H$\beta$ emission lines (plus \oiii) in SDSS~J0842$+$3625 (see Table~1 in Zakamska et al. 2005). \\
$\dagger$ For this source, the spectral parameters were obtained using a scattering model (with $\approx$~3~per~cent 
of scattered component), similarly to P06;  
the \xray\ luminosity was corrected to account only for the nuclear emission. 
\end{minipage}
\end{table*}

\subsection{X-ray spectral analysis: SDSS~J0050$-$0039 and SDSS~J1153$+$0326}

Here we show the \xray\ spectral results for the two targets with the most 
counts in the \chandra\ images analysed in this paper, SDSS~J0050$-$0039 
($\approx$~41 full-band counts) and SDSS~J1153$+$0326 
($\approx$~80 full-band counts; see Table~\ref{chandra_counts}). 
We look at the properties of the other sources in $\S$3. 

\begin{figure*}
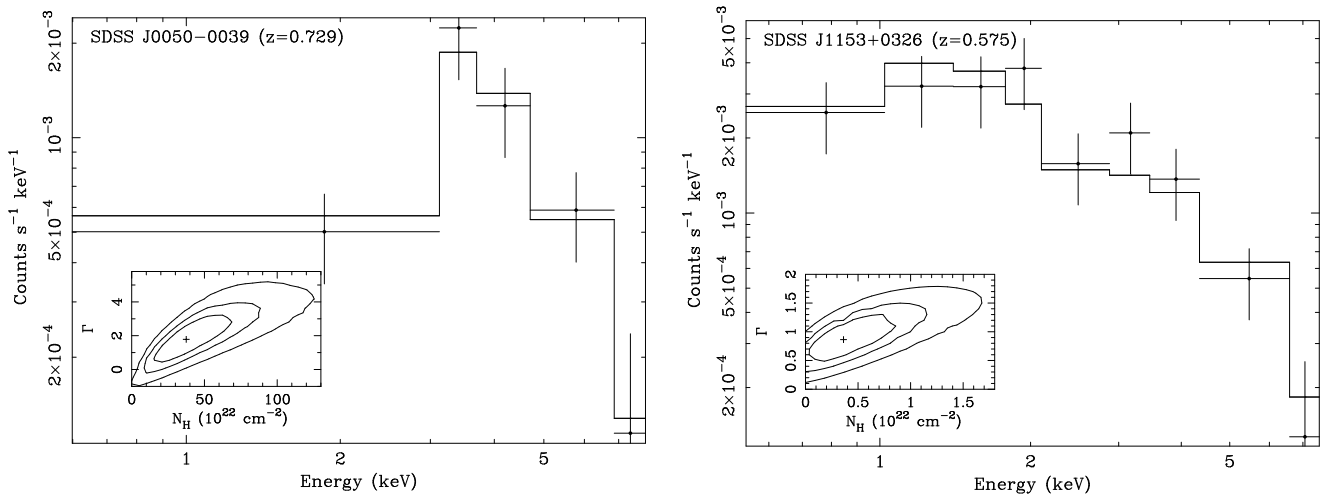

\vglue3.6cm
\hglue-3.3cm
\includegraphics[angle=-90,width=0.18\textwidth]{vignali.fig3a.ps}
\hskip5.7cm
\includegraphics[angle=-90,width=0.18\textwidth]{vignali.fig3b.ps}
\vskip0.9cm
\caption{
\xray\ spectra of SDSS~J0050$-$0039 (left panel) and SDSS~J1153$+$0326 
(right panel) fitted with an absorbed power-law model and the $C$-statistic; 
the data were rebinned for presentation purposes. 
The 68, 90, and 99~per~cent confidence contours for the column density and 
photon index are shown in the insets (see $\S$2.3 for details). 
}
\label{new_spectra}
\end{figure*}

Spectral analysis was carried out with {\sc xspec} Version 11.3.2 
(Arnaud 1996) using unbinned data and the $C$-statistic (Cash 1979);  
given the low number of counts, this statistical approach is the most viable 
to provide basic constraints on the \xray\ emission of these sources 
(e.g., Nousek and Shue 1989) without losing spectral information (as, e.g., 
in the case of the hardness-ratio or band-ratio analyses). 
Errors are quoted at the 90~per~cent confidence level for one 
interesting parameter (\hbox{$\Delta$$C=2.71$}; Avni 1976; Cash 1979), 
unless stated otherwise. 
Initially, the two \xray\ spectra are fitted with a power law and 
Galactic absorption only (from Dickey and Lockman 1990; see Table~\ref{xray_param}). 
In both cases, the flat photon index of the power-law component 
(\hbox{$\Gamma=-0.78^{+0.50}_{-0.55}$} for SDSS~J0050$-$0039 and 
\hbox{$\Gamma=0.56\pm{0.28}$} for SDSS~J1153$+$0326) 
is suggestive of absorption, which is therefore included in the spectral fitting. 
Intrinsic absorption is statistically significant 
for SDSS~J0050$-$0039 if $\Gamma$ is left free to vary 
\hbox{[$N_{\rm H}(z=0.729)=3.75^{+3.12}_{-2.41}\times10^{23}$~cm$^{-2}$}, with 
\hbox{$\Gamma=1.78^{+1.22}_{-1.41}$}; see Fig.~\ref{new_spectra}],\footnote{The column density reported here 
is slightly different from that shown in Table~\ref{xray_param}, since the latter is obtained with $\Gamma=2.0$.} 
while for SDSS~J1153$+$0326 the implied absorption is 
\hbox{$N_{\rm H}(z=0.575)=1.54^{+0.90}_{-0.54}\times10^{22}$~cm$^{-2}$} for a fixed photon index 
of $\Gamma=2.0$ (as observed in local and high-redshift AGN; e.g., 
Reeves \& Turner 2000; Piconcelli et al. 2005; Page et al. 2005; Shemmer 
et al. 2005, 2006). 
Given the presence of obscuration and the high de-absorbed, rest-frame 
\hbox{2--10~keV} luminosities (7.2$\times10^{44}$~\lum\ and 1.2$\times10^{44}$~\lum\ for 
SDSS~J0050$-$0039 and SDSS~J1153$+$0326, respectively) of these two sources, 
we conclude that both the \xray\ and optical spectra indicate that these objects are Type~2 quasars. 
%

\section{The ``extended'' sample: the properties of SDSS Type~2 quasars}

For our analyses of the overall properties of the SDSS Type~2 quasars, we extend 
the main ($\S$2.1) and serendipitous ($\S2.2$) samples with eight archival 
\chandra\ and \xmm\ observations (open squares in Fig.~\ref{lxoiiiz}); 
we remind that for the sources observed by \rosat\ (see $\S$1), a detailed analysis is 
presented in V04 and Vignali et al. (2004b), while the \xray\ spectra of 
the sources with enough counts for \xray\ fitting have been shown already 
in V04 and P06.
Although the sources shown in this section (and summarized in Table~\ref{obs_log}) 
have already been published (V04; P06), we preferred to 
retrieve all the available \xray\ data from the archives and perform a uniform 
analysis by adopting the same reduction techniques described in this 
paper (for the \chandra\ data) and in V04 (for the \xmm\ data). 
We note that the agreement between our results and those of P06 
is generally good. 

We extracted the \xray\ counts in the observed \hbox{0.5--8~keV} energy range 
for \chandra\ (see Table~\ref{chandra_counts}) and in the observed \hbox{0.5--10~keV} band 
for \xmm\ (Table~\ref{xmm_counts}). For off-axis sources, we took into account the 
broadening of the PSF at the off-axis angles corresponding to the source 
position and, in the computation of the count rates reported in 
Tables~\ref{chandra_counts} and \ref{xmm_counts}, we adopted the exposure times derived from the exposure 
maps. At large off-axis angles (reported in Table~\ref{obs_log}), the difference with 
respect to the ``nominal'' exposure times can be significant (up to $\approx$~50~per~cent). 
%
\begin{table}
\begin{minipage}{75mm}
\caption{X-ray counts: archival \xmm\ observations.}
\label{xmm_counts}
\footnotesize
\begin{tabular}{lrrrr}
\hline
   & \multicolumn{3}{c}{X-ray counts} & Count~rate \\
\cline{2-4} \\
   & pn & MOS1 & MOS2 & pn \\
SDSS~J0210$-$1001 & 154   & 59    & 62    & 1.93$\pm{0.16}$ \\
SDSS~J0243$+$0006 & $<29$ & $<22$ & $<18$ & $<0.20$ \\ 
SDSS~J1226$+$0131 & 267   & 98    & 120   & 3.43$\pm{0.22}$ \\
SDSS~J1641$+$3858 & 963   & 411   & 412   & 7.89$\pm{0.27}$ \\
\hline
\end{tabular}
\end{minipage}
\begin{minipage}{75mm}
Counts and count rates are computed in the \hbox{0.5--10~keV} band; 
count rates are in units of $10^{-2}$ counts~s$^{-1}$. 
\end{minipage}
\end{table}

%

\subsection{Obscured and Compton-thick quasars}

To draw some first results for the Type~2 quasars in the ``extended'' sample of 
16 sources (Table~\ref{xray_param}), we compared their rest-frame, de-absorbed 
\hbox{2--10~keV} luminosity 
(when a direct measurement of the column density from the \xray\ spectrum is possible, see 
Fig.~\ref{complum}a; in the other cases, it is the measured \xray\ luminosity, see Fig.~\ref{complum}b) 
with the \xray\ luminosity range 
predicted by the Mulchaey et al. (1994) correlation (see $\S$2.1). 

\begin{figure}
\includegraphics[angle=0,width=85mm]{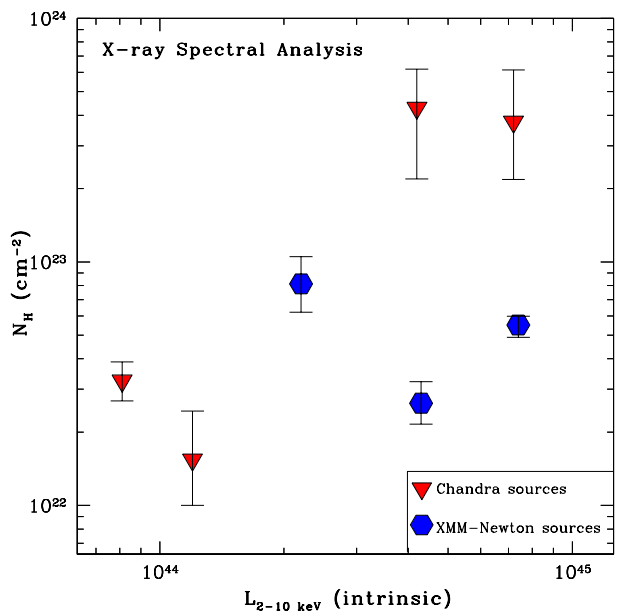}
\caption{
Intrinsic (de-absorbed) \hbox{2--10~keV} luminosity vs. 
column density for the seven sources in the ``extended'' sample with enough counts 
for \xray\ spectral analysis. The error bars represent the 90~per~cent confidence 
level for the column density measurement. 
}
\label{lxnh}
\end{figure}

\begin{figure*}
\includegraphics[angle=0,width=85mm]{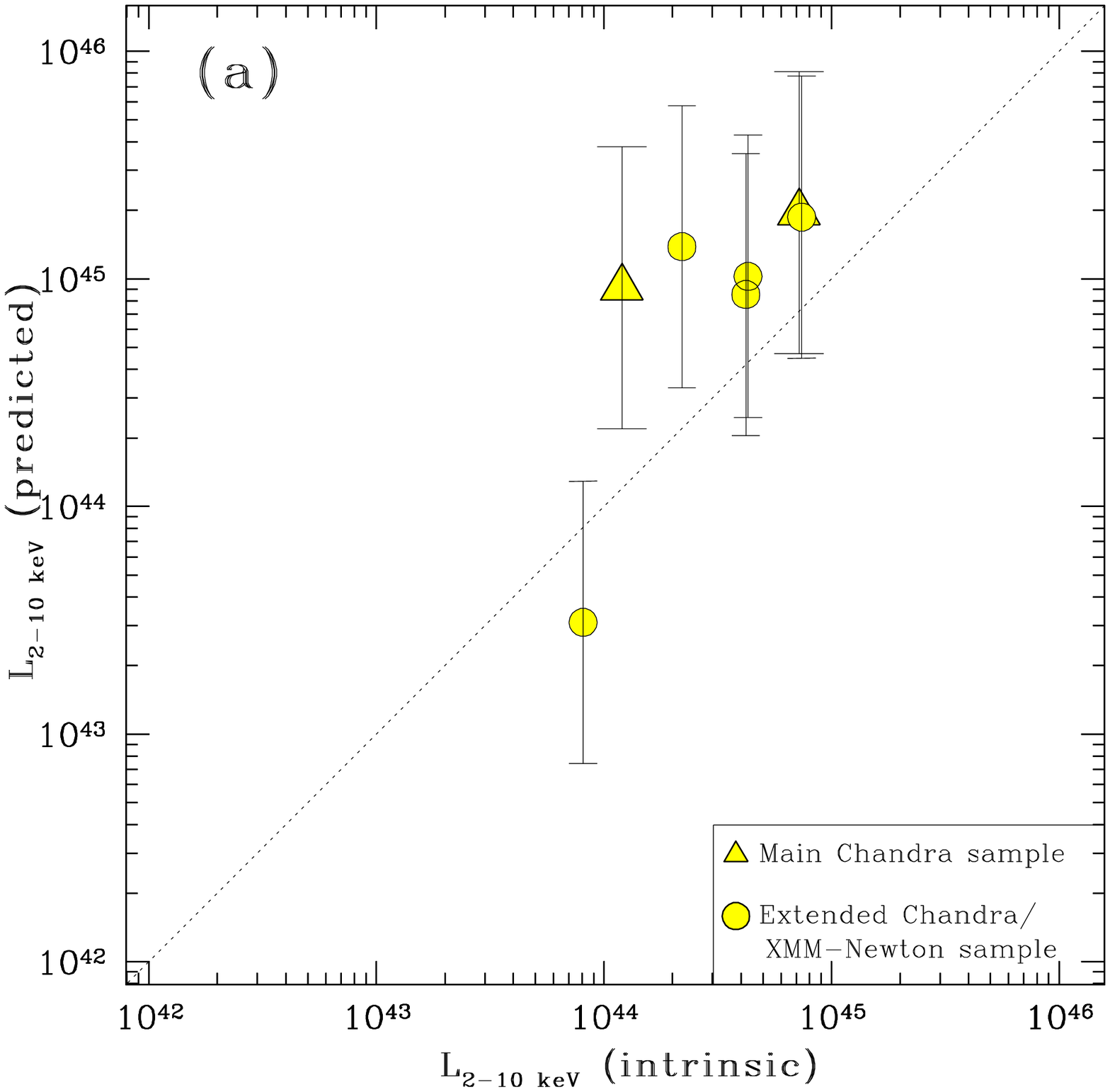}
\includegraphics[angle=0,width=85mm]{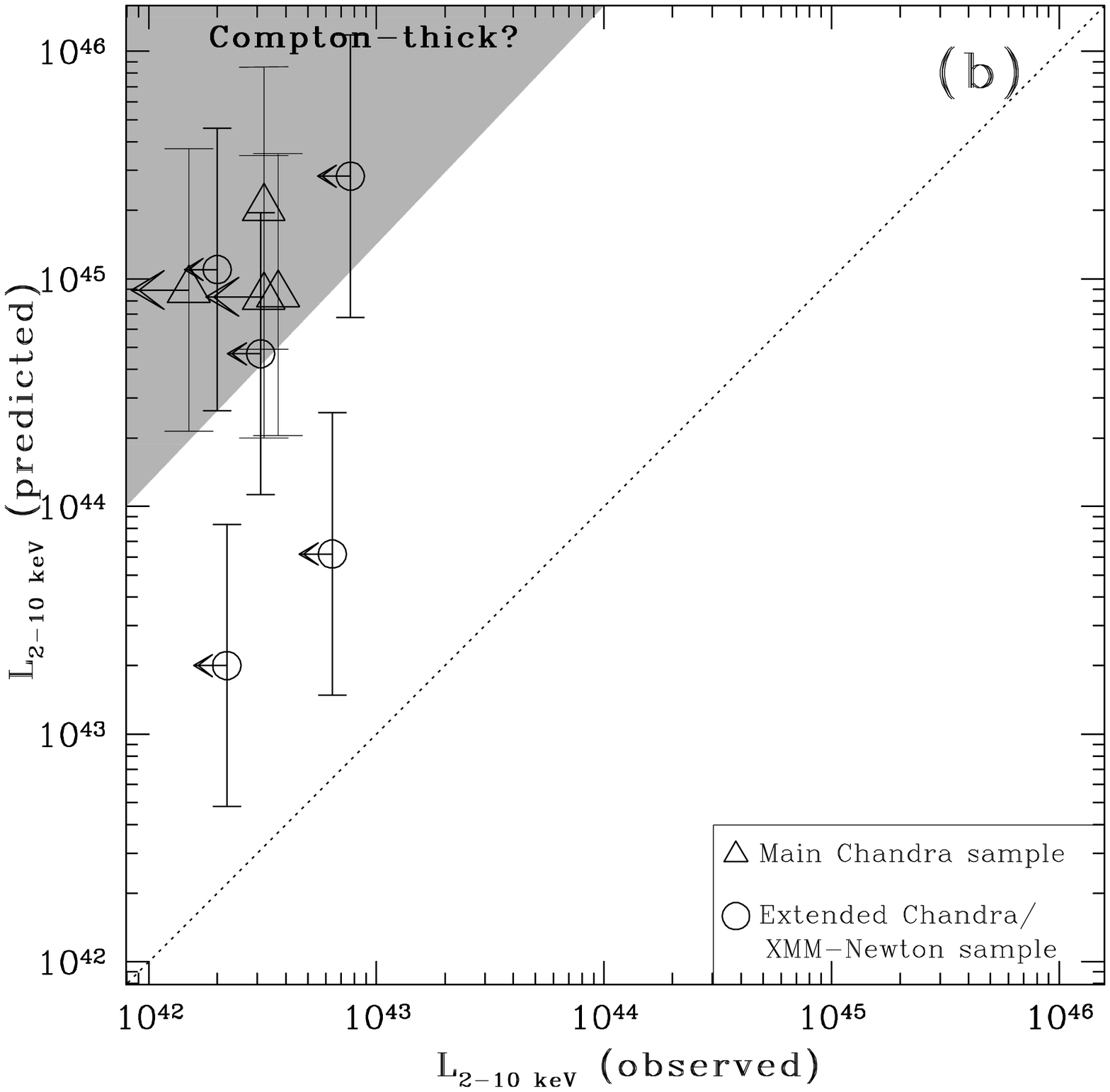}
\caption{
Comparison of the \hbox{2--10~keV} luminosity computed from the available \xray\ data 
with that predicted assuming the Mulchaey et al. (1994) correlation. 
The dotted line shows the 1:1 ratio between the two \xray\ luminosities. 
(a) Sources whose \xray\ luminosity is de-absorbed on the basis of the column density 
($N_{\rm H}$ up to $\approx5\times10^{23}$~cm$^{-2}$), 
computed directly from the \xray\ spectra (see Table~\ref{xray_param}). 
(b) All of the remaining sources, for which the \xray\ luminosity is derived from the \xray\ flux 
with no correction for the unknown column density. 
Left-ward arrows indicate upper limits on the observed \xray\ luminosity, while 
the grey region shows the locus of extremely obscured, likely 
Compton-thick ($\gtrsim10^{24}$~cm$^{-2}$) objects, where the observed luminosity is 
less than 1~per~cent of the predicted one. 
}
\label{complum}
\end{figure*}

All of the sources with enough counts for a basic or moderate-quality 
\xray\ spectral analysis (seven in the current sample) are obscured, 
with column densities in the range \hbox{$\approx$~10$^{22}$ -- a~few~10$^{23}$~cm$^{-2}$} 
(see Table~\ref{xray_param} and Fig.~\ref{lxnh}). For two of these sources 
(the ones whose error bars do not touch the 1:1 dotted line in Fig.~\ref{complum}a), the de-absorbed 
\hbox{2--10~keV} luminosities are below the lower bound of the predicted \xray\ luminosity range. 
%
%
This might suggest that either absorption was not computed properly due to the limited photon 
statistics or the predictions from the \oiii\ line vs. hard \xray\ flux correlation 
sometimes over-estimates the nuclear flux (see Netzer et al. 2006 for a discussion). 
We also point out that because of the limited wavelength coverage of the SDSS spectroscopy 
($\approx$~2450--5900\AA\ in the source rest frame at the average redshift of 0.56) 
it was not possible to estimate the reddening via the H$\alpha$/H$\beta$ Balmer decrement 
and therefore no corrections were applied to the \oiii\ flux to account for the absorption 
due the narrow-line region itself (for details, see Maiolino et al. 1998; Bassani et al. 1999 
and the recent results presented by Panessa et al. 2006); it is therefore possible that the prediction of the \xray\ 
luminosity would be higher for at least some sources. 
A more detailed analysis on all these issues will be carried out when a larger sample of SDSS 
Type~2 quasars with good-quality \xray\ spectra are available. 
%
%

\begin{figure}
\includegraphics[angle=0,width=85mm]{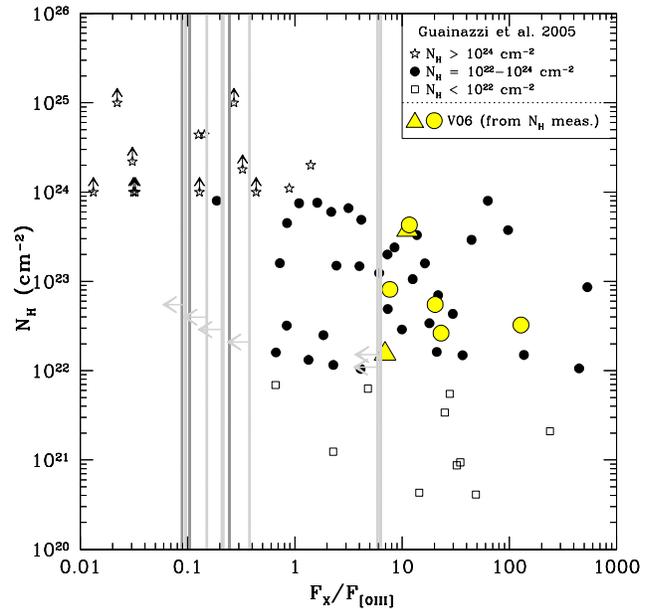}
\caption{
\hbox{2--10~keV}-to-\oiii\ flux ratio vs. intrinsic column density for the sources 
presented in this paper (large filled triangles and circles, as in 
Fig.~\ref{complum}a) 
and the Seyfert~2 galaxies from Bassani et al. (1999) and Guainazzi et al. (2005). 
The \xray\ fluxes are not corrected for absorption. 
Dark (light) grey regions indicate the X-ray-to-\oiii\ flux ratios for the sources 
with \xray\ detections (upper limits, with leftward-pointing arrows) 
with no direct column density measurements. 
}
\label{fxoiii}
\end{figure}

The situation appears different for the \xray\ undetected sources and for those with 
a limited number of counts (Fig.~\ref{complum}b). 
All of these sources have observed (i.e., not corrected for absorption) \hbox{2--10~keV} 
luminosities generally far below (up to more than two orders of magnitude) 
the predicted \xray\ luminosity range (see Table~\ref{xray_param}). 
On the basis of their optical spectra, there is no evidence that these sources host 
active nuclei of low luminosity and therefore the \xray\ observations suggest that 
these sources are even more obscured than those for which \xray\ absorption has been derived 
from direct \xray\ spectral fitting. 
It is possible that the \xray\ brightest sources are the ``tip of the iceberg'' of 
the SDSS Type~2 quasars, with $N_{\rm H}\approx$~10$^{22}$ -- a~few~10$^{23}$~cm$^{-2}$ (Fig.~\ref{complum}a) 
and a significant fraction (up to $\approx$~50~per~cent) of the population is 
characterized by column densities $\gg10^{23}$~cm$^{-2}$, with many Compton-thick 
($N_{\rm H}\gtrsim10^{24}$~cm$^{-2}$) AGN (Fig.~\ref{complum}b; see P06 for similar conclusions). 
As further support to this indication, in Fig.~\ref{fxoiii} we plot the \xray-to-\oiii\ flux ratio 
vs. column density for the 16 SDSS Type~2 quasars and the Seyfert~2 galaxies from 
the Bassani et al. (1999) and Guainazzi et al. (2005) compilations. 
%
%
While the SDSS Type~2 quasar candidates with direct column density measurements 
(filled triangles and circles) are in the obscured AGN region, most of the sources 
with lower photon statistics or undetected in the \xray\ band (marked by dark and light-grey 
vertical lines, respectively, the latter with leftward-pointing arrows) lie in the 
$F_{\rm X}/F_{[OIII]}$ region which is mostly populated by Compton-thick AGN. 
Provided that these results will be confirmed by further observations, there is evidence that 
the optical selection criteria adopted by Zakamska et al. (2003) on the basis of the SDSS spectra 
are highly effective in picking up moderate-redshift Compton-thick AGN.

\subsection{Comparison with ROSAT results}

Five sources are in common between the sample observed by \rosat\ (V04) 
and the current ``extended" sample (plotted as squared circles in Fig.~\ref{lxoiiiz}): 
besides SDSS~J1226$+$0131, whose \xmm\ 
observation and spectrum were originally presented by V04, we find 
SDSS~J0842$+$3625 (\chandra\ upper limit), SDSS~J0210$-$1001 and 
SDSS~J1641$+$3858 (both \xmm\ detections with \xray\ spectra available), 
and SDSS~J0243$+$0006 (being undetected by \xmm\ at a large off-axis angle). 
We note that the soft (\hbox{0.5--2~keV}) flux measurements obtained by 
the observations presented in this paper are deeper than those, relatively shallow, 
obtained by \rosat, and are generally in agreement, with the 
exception of SDSS~J1641$+$3858: for this source, the soft \xray\ flux 
measured by \xmm\ is a factor of $\approx$~2 higher than the 3$\sigma$ upper 
limit reported in V04. 
We note that it is possible that variability either in the nuclear/scattered 
flux or in the absorber might have occurred in the 
time interval between the \rosat\ and \xmm\ observations ($\approx$~11 years 
in the observed frame). 

For all of these sources, the \rosat\ analysis suggested the presence of 
absorption in the \xray\ band (see Table~2 in V04), and the column densities 
obtained from \xray\ spectral fitting (Table~\ref{xray_param}) are generally consistent, 
within the uncertainties, with those derived using \rosat\ data. 

%
\begin{table}
\begin{minipage}{75mm}
\caption{X-ray detections and upper limits vs. on-axis/off-axis observations.}
\label{summary}
\footnotesize
\begin{tabular}{ccccccccc}
\hline
\multicolumn{4}{c}{On-axis observations} & & \multicolumn{4}{c}{Off-axis observations} \\
\cline{1-4} \cline{6-9} \\
\multicolumn{2}{c}{\chandra} & \multicolumn{2}{c}{\xmm} & & \multicolumn{2}{c}{\chandra} & \multicolumn{2}{c}{\xmm} \\
\multicolumn{1}{c}{det} & \multicolumn{1}{c}{ul} & \multicolumn{1}{c}{det} & \multicolumn{1}{c}{ul} & & 
\multicolumn{1}{c}{det} & \multicolumn{1}{c}{ul} & \multicolumn{1}{c}{det} & \multicolumn{1}{c}{ul} \\
6(3) & 2 & 2(2) & 0 & & 1(1) & 3 & 1(1) & 1 \\
\hline
\end{tabular}
\end{minipage}
\begin{minipage}{75mm}
``det'' means \xray\ detection, while ``ul'' indicates an \xray\ upper limit. 
The number of sources with available X-ray spectra (seven in the whole sample) 
is reported between parenthesis. 
\end{minipage}
\end{table}

%

\subsection{Estimate of the number density of SDSS Type~2 quasars}

Overall, it appears clear that the SDSS is able to provide a large sample of 
optically selected Type~2 quasars which are confirmed as such by \xray\ 
observations and, possibly, a significant number of candidate Compton-thick AGN. 
To estimate the role of these Type~2 quasars in the 
AGN synthesis models for the XRB, it is compelling to derive their surface 
density and luminosity function and to extend the \xray\ spectral analysis 
carried out in this paper to a larger sample. 
At present, given the highly inhomogeneous selection and incompleteness of the sample of 291 
Type~2 AGN candidates presented by Zakamska et al. (2003), the number density of these objects 
is subject to large uncertainties and has not been accurately calculated yet. 
However, we can provide a first-order estimation. If we consider that 48.5 per cent of the plates 
used by Zakamska et al. (2003) were taken from the SDSS Data Release~1 (Abazajian et al. 2003), 
whose spectroscopic coverage is 1360 deg$^{2}$, and assume that the SDSS Type 2 AGN are 
uniformly distributed, 
then their surface density would be 
$\approx$~0.1~deg$^{-2}$. Since only approximately half of the sample can be associated 
with high-luminosity AGN on the basis of the \oiii\ luminosity (see $\S$6.2 of 
Zakamska et al. 2003), the resulting estimate of the surface density of SDSS Type~2 quasars is 
$\approx$~0.05~deg$^{-2}$. 
At the average \hbox{2--10~keV} flux of the seven luminous obscured sources 
(from \xray\ spectral fitting) in our sample ($\approx$~1.8$\times10^{-13}$~\cgs), 
we expect a surface density of $\approx$~0.5~deg$^{-2}$ for Type~2 quasars from the La Franca 
et al. (2005) luminosity-dependent density evolution model (also see Cocchia et al., submitted). 
If we limit our study to the redshift range $\approx$~0.3--0.8 of the Zakamska et al. (2003) 
sample, then the surface density of Type~2 quasars becomes $\approx$~0.15~deg$^{-2}$ 
(R. Gilli, private communication; see also Gilli et al. 2006). 
Although our estimate of the surface density of SDSS Type~2 quasars is affected 
by significant uncertainties, 
it demonstrates that \xray\ observations probably provide a more complete 
census of obscured quasar activity than that achieved at optical wavelengths.
We note, in fact, that a sizable number of \xray\ selected Type~2 quasars 
is not characterized by strong narrow emission lines (e.g., Mignoli et al. 2004).

\section{Summary of the results on SDSS Type~2 quasars}

In this paper we have reported new \chandra\ exploratory 
observations of six SDSS Type~2 quasar 
candidates at \hbox{$z$=0.49--0.73} (which constitute the so-called ``main'' 
sample), selected among the most \oiii\ luminous 
objects from the Zakamska et al. (2003) sample, and provide results on 
two additional Type~2 quasar candidates serendipitously fallen in the same observations. 
We have also re-analysed archival \chandra\ and \xmm\ observations 
of eight additional SDSS Type2 quasar candidates which constitute, along with 
the above sources, the ``extended'' sample (for a total of 16 sources), 
thus providing a progress report 
on the \xray\ properties of SDSS Type~2 quasars with respect to the preliminary results 
presented by V04. 
Although the strategy adopted to gain the final sample of 16 SDSS Type~2 
quasars is quite varied, from Table~\ref{summary} it appears evident that \chandra\ 
exploratory ($<$10~ks) on-axis observations are able to provide a significant fraction of 
\xray\ detections (6/8, 75~per cent), but the 
fraction of sources with \xray\ spectral information is clearly limited (3/6, 50~per~cent). 
On the other side, 
the high-energy throughput allowed by the EPIC cameras on-board \xmm\ guarantees, 
for obscured sources at relatively bright \xray\ fluxes, 
good-quality \xray\ spectra, at least at small off-axis angles (Table~\ref{summary}). 
The principal results can be summarized as follows: 
\begin{description}
\item[$\bullet$] 
Ten of the ``extended'' sample of 16 SDSS Type2 quasar candidates 
were detected either by \chandra\ or \xmm\ (see Table~\ref{summary}). 
For seven of these sources, basic/moderate-quality \xray\ spectral analysis has 
allowed us to achieve a direct measurement of the absorption 
(\hbox{$\approx$~10$^{22}$ -- a~few~10$^{23}$~cm$^{-2}$}). 
These results are consistent with those obtained using \rosat\ data and reported by V04. 
\item[$\bullet$] 
Using the \oiii\ line flux as an indicator of the intrinsic \xray\ emission, 
as verified for the seven sources with higher \xray\ statistics, 
we obtained indications that the \xray\ 
undetected sources and the sources with a limited number of counts are possibly 
more obscured than those found absorbed through direct \xray\ spectral fitting. 
This would imply that up to $\approx$~50~per~cent of the population 
is characterized by column densities $\gg10^{23}$~cm$^{-2}$, with a sizable number 
of Compton-thick quasars possibly hiding among the \xray\ faintest sources. 
This possibility is also suggested by the comparison of the \xray-to-\oiii\ 
flux ratios of our sources vs. those obtained from a large sample of 
Seyfert~2 galaxies. 

\end{description}

\section*{Acknowledgments}
CV and AC acknowledge partial support by the Italian Space agency 
under the contract ASI--INAF I/023/05/0. 
DMA is supported by the Royal Society. 
The authors would like to thank R. Gilli, M. Mignoli and G. Zamorani for useful discussions, 
M. Guainazzi for providing data for Fig.~\ref{fxoiii}, and the referee for his/her 
suggestions which helped us to improve the manuscript. 

\end{document}